\documentclass[preprint,showpacs,showkeys,preprintnumbers]{revtex4}
\usepackage{amsmath}
\usepackage{amssymb}

\begin{document}

\title{Quantum Optical Random Walk: Quantization Rules and Quantum
Simulation of Asymptotics }
\author{Demosthenes Ellinas $^\dag$ and Ioannis Smyrnakis $^\ddag$ }

\begin{abstract}
Rules for quantizing the walker+coin parts of a classical random walk are
provided by treating them as interacting quantum systems. A quantum optical
random walk (QORW), is introduced by means of a new rule that treats quantum
or classical noise affecting the coin's state, as sources of quantization.
The long term asymptotic statistics of QORW walker's position that shows
enhanced diffusion rates as compared to classical case, is exactly solved. A
quantum optical cavity implementation of the walk provides the framework for
quantum simulation of its asymptotic statistics. The simulation utilizes
interacting two-level atoms and/or laser randomly pulsating fields with
fluctuating parameters.
\end{abstract}



%
%

\address{
Technical University of Crete, Department of Sciences\\
 Division of Mathematics, GR 731 00, Chania, Greece.\\
\ \ e-mail: $^\dag$ \texttt{ellinas@science.tuc.gr} \ \ $^\ddag$
\texttt{smyrnaki@tem.uoc.gr} } 
\maketitle

\section{Introduction}

Discrete quantum random walks on a line, \cite{aharonov},\cite{meyer},\cite%
{aa},\cite{ambainis}, (for a review in a more general setting see \cite%
{kempe}) is a paradigmatic construction of a quantum system performing a
motion similar to the usual classical discrete random walk on a line, but
with a number of important differences like the quantum treatment of its
coin and walker systems, the role of quantum entanglement, novel diffusion
and hitting properties of its motion etc. The asymptotic behavior of such
walks has been studied in \cite{ambainis},\cite{konno},\cite{grimmett1}.
Also important theoretical suggestions about the utilization of those
properties in the construction of quantum algorithms that would outperform
classical rivals in tasks such as searching a database etc, have been put
forward, see e.g. \cite{kempe1}. The effect of decoherence on the evolution
of discrete quantum walks has been studied in e.g. \cite{brun},\cite{kendon}%
. \ Experimental proposals also exists concerning e.g. quantum coin-tossing,
and quantum diffusion see e.g. \cite{travaglione},\cite{zeil},\cite{dur},%
\cite{sanders}, \cite{ryan}. The coin control over discrete quantum walks on
graphs has been studied in \cite{trag}. Continuous time quantum walks on
graphs, which are defined without the use of a coin, were studied in e.g.
\cite{fahri2},\cite{childs}, and their potential implementation for the
construction of fast search algorithms were studied in \cite{childs1}.

The present work extends previous ones on the so called $V^{k}$ models of
quantum discrete walks on a line\cite{ellinas1}\cite{ellinas}\cite{ellsmyrn1}%
\cite{ellsmyrn2}. Its aim is twofold: first to argue that a general and
systematic framework of quantization of classical walks is possible and well
physically motivated, and second to study the walk beyond transient dynamics
in its asymptotic regime and to show the possibility of simulating its
statistical behavior in terms of another quantum system, introducing thus
the quantum simulation of a quantum walk. This double aim is achieved by
introducing the Quantum Optical Random Walk (QORW); a mathematical model
that refers to a typical cavity QED type of arrangement where a beam of
two-level atoms passes through an optical cavity. [Technical caveat:
although the mathematical description of cavity field done here is not the
usual one i.e. in terms of the boson degree of freedom, but instead is in
terms of states with positive and negative valued energy also (canonical
algebra vs. Euclidean algebra, see below), appropriate truncations can be
introduced in order to have only positive energy states \textbf{(}c.f. e.g.
\cite{sanders}\textbf{)}; this remedy for the problem however will not be
discussed here. ]

The outline of the paper goes as follows: in Chapter II, the study of
quantization rules is developed, the QORW is introduce in Chapter III and
several example of walks are solved; Chapter IV treats the walk in its
asymptotic regime and introduces the technique of quantum simulation,
examples of numerically simulated asymptotic probability density functions
are provided; conclusions are summarized in last Chapter V.

\section{Quantization Rules for Classical Random Walks}

The essential feature of a simple QRW on a line is the promotion of
mathematical correspondence: \textit{left/right}$\rightarrow $\textit{\
head/tails, }between \textit{walker}'s move directions and \textit{coin}'s
two sides, to a dynamical interaction among two physical systems. This is
realized by introducing state Hilbert spaces $H_{w}=span(|m\rangle )_{m\in Z}
$ and $H_{c}=span(|+\rangle ,|-\rangle ),$ for quantum walker and coin
systems respectively. In $H_{w}$ operates the Euclidean algebra with
generators the \textit{step operators} $E_{\pm }|m\rangle =|m\pm 1\rangle ,$%
(and their their powers e.g. for $a>0,$ $(E_{\pm })^{a}|m\rangle \equiv
E_{\pm a}|m\rangle =|m\pm a\rangle ),$ and the \textit{position operator }$%
L|m\rangle =m|m\rangle .$ They satisfy the commutation relations $[L,E_{\pm
}]=\pm E_{\pm },$ $[E_{+},E_{-}]=0.$ Also important is the Fourier basis $%
H_{w}=span(|\phi \rangle =\frac{1}{2\pi }\sum_{m\in Z}e^{im\phi }|m\rangle
,0\leq \phi <2\pi ),$ which is the eigenbasis of step operators viz. $E_{\pm
}|\phi \rangle =e^{i\phi }|\phi \rangle .$ In the coin space the projection
operators $P_{\pm }=|\pm \rangle \langle \pm |,$ are needed in order to
realize the coin-tossing that drives the walk. Indeed one step of classical
random walk (CRW), is described by means of the unitary $V_{cl}=P_{+}\otimes
E_{+}+P_{-}\otimes E_{-}.$ Its conditional action on walker states realizes
the coin-tossing and the subsequent move of walker.

Explicitly the coin $\rho _{c}$\ and walker $\rho _{w},$\ density matrices
initially taken in product form $\rho _{c}\otimes \rho _{w},$ are assumed to
interact unitarily by the transformation $\rho _{c}\otimes \rho
_{w}\rightarrow V_{cl}(\rho _{c}\otimes \rho _{w})V_{cl}.$Subsequently the
two systems are decoupled by an unconditional measurement of the coin
subsystem realized by means of the partial trace i.e. $V_{cl}(\rho
_{c}\otimes \rho _{w})V_{cl}\rightarrow Tr_{c}(V_{cl}(\rho _{c}\otimes \rho
_{w})V_{cl})$\ ; the latter constitues the dynamical realization of the coin
tossing process. The resulting walker density matrix $\varepsilon
_{V_{cl}}(\rho _{w})=Tr_{c}(V_{cl}(\rho _{c}\otimes \rho _{w})V_{cl}),$%
written in the eigenbasis states of position operator $L$\ (see above),
provides by means of its diagonal elements the occupation probability
distribution $p_{m}=\langle m|\varepsilon _{V_{cl}}(\rho _{w})|m\rangle ,$\
of the states of walker system. This distribution in the course of time
steps of the walk $n=1,2,...$, is identified with the occupation
probabilities of the classical random walk on integers i.e. $%
p_{m}^{(n)}=\langle m|\varepsilon _{V_{cl}}(\rho _{w}^{(n)})|m\rangle
,m=0,\pm 1,\pm 2,...,$with bias determined by the elements of $\rho _{c}$.
Due to this the $\varepsilon _{V_{cl}}$\ map is conceived as realization of
CRW, which we next seek to quantize.

Quantization of CRW is conceived as the incorporation in coin space of an
additional unitary operator $U,$ the coin reshuffling matrix, so that the
one-step operator now becomes $V=V_{cl}U\otimes \mathbf{1.}$ To facilitate
conceptual comparisons this procedure was christened $U-$quantization in
ref. \cite{ellsmyrn2}. Almost all work that has been done in the area of
quantum random walks, has been based on the scheme of $U-$quantization or
modifications thereof. One particular class of such $U-$quantized walks are
the $V^{k}-$models for which the one-step evolution of the walker's density
matrix is given by the CPTP map $\varepsilon _{V^{k}}$ as $\varepsilon
_{V^{k}}(\rho _{w})=Tr_{c}V^{k}(\rho _{c}\otimes \rho _{w})V^{\dagger k}$%
\cite{vk}.



Next we introduce a generalized version of the previous quantization method,
the $\varepsilon -$quantization rule, which employs a positive and
completely positive trace preserving map $\varepsilon ,$ acting on the coin
density matrices, which is not necessarily taken to be unital, namely $%
\varepsilon (\frac{1}{2}\mathbf{1})\neq \frac{1}{2}\mathbf{1.}$ Then the
one-step evolution of walker's density matrix of an $\varepsilon -$quantized
model of a classical walk is defined as
\begin{equation}
\varepsilon _{V}(\rho _{w})=Tr_{c}V_{cl}\varepsilon \otimes \mathbf{1}(\rho
_{c}\otimes \rho _{w})V_{cl}^{\dagger }.
\end{equation}%
In general for a $V^{k}$ quantum walk model we will have that
\begin{equation}
\varepsilon _{V^{k}}(\rho _{w})=Tr_{c}\left[ V_{cl}\varepsilon _{k}\otimes
\mathbf{1...}\left( V_{cl}\varepsilon _{1}\otimes \mathbf{1}(\rho
_{c}\otimes \rho _{w})V_{cl}^{\dagger }\right) ...V_{cl}^{\dagger }\right] ,
\end{equation}%
where in general a different quantizing CP map $\varepsilon $ can be used
between coin+walker interactions. To appreciate the changes brought about by
$\varepsilon -$quantization we use the adjoint action of e.g operator $X,$
on a density matrix defined as $AdX(\rho )=X\rho X^{\dagger },$with property
$AdXY(\rho )=AdXAdY(\rho ).$ We assume that the $\varepsilon $ map is
determined by a set of Kraus operators as $\varepsilon (\rho
_{c})=\sum_{i}S_{i}\rho _{c}S_{i}^{\dagger },$ or in terms of the adjoint
action $\varepsilon (\rho _{c})=\sum_{i}AdS_{i}(\rho _{c}),$\cite%
{nielsenchuang}. To contrast deference between the usual $U-$quantization
and the proposed $\varepsilon -$quantization rules we write the evolution
map of the former as
\begin{eqnarray}
\varepsilon _{V^{k}}(\rho _{w}) &=&Tr_{c}\left( AdV_{cl}AdU\right) ^{k}(\rho
_{c}\otimes \rho _{w})  \notag \\
&=&Tr_{c}\left( Ad(V_{cl}U)\right) ^{k}(\rho _{c}\otimes \rho _{w}),
\end{eqnarray}%
and the evolution map of the latter as
\begin{eqnarray}
\varepsilon _{V^{k}}(\rho _{w}) &=&Tr_{c}\left(
\sum_{i}AdV_{cl}AdS_{i}\right) ^{k}(\rho _{c}\otimes \rho _{w})  \notag \\
&=&Tr_{c}\left( \sum_{i}Ad(V_{cl}S_{i})\right) ^{k}(\rho _{c}\otimes \rho
_{w}).
\end{eqnarray}

It is evident from above that the $\varepsilon -$quantization rule applies a
sum of adjoints at each factor of the $k-$fold product. This sum in addition
to $V_{cl},$ is determined by the Kraus generators of the quantization map $%
\varepsilon .$ In the simplest case where there is only one single unitary
Kraus generator, the $\varepsilon -$ rule reduces to the $U-$quantization
rule. The inclusion of Kraus generators in the quantization of \ classical
walk, may stem from the fact of a hidden quantum interaction between the
coin system and another unobserved quantum system, or it may be due to some
classically fluctuating parametric variance of the coin system, that is
treated on the average; generically we may say that it is due to quantum or
classical noise. (see also previous studies of non-unitary models \cite%
{travaglione,sanders,dur,lp,ks,rom}, The number of Kraus generators as well
as their matrix type provide in any case a flexible framework in which the
quantization map may incorporate complex physical processes in coin systems
that may take place within a evolution step of the walk.

\section{Quantum Optical Random Walk}

As an application of the $\varepsilon -$quantization rule we introduce a new
kind of quantum walk. To this end we need to introduce a continuous family
CPTP maps $\mathbf{E}=\{t\rightarrow \varepsilon _{t},t\geq 0\},$ acting on
the space of coin density matrices, where variable $t$ is taken to stand for
time. Among members of family $\mathbf{E}\emph{,}$ there exists a semi-group
composition law, i.e $\varepsilon _{t_{1}}\circ \varepsilon
_{t_{2}}=\varepsilon _{t_{1}+t_{2}}.$Also the identity map $\varepsilon
_{0}=id,$ is included in $\mathbf{E}\emph{.}$

Now we come to the Quantum Optical Random Walk; it physically outlines the
crossing of an beam of two-level atoms, the \textit{coins, }through a
quantum optical cavity which sustains a standing quantum mode identified
with the \textit{walker }system. The walker+coin interaction realizes a $%
V^{2}$ QRW model, and it also takes into account the interaction of coin
with some external environment, formalized as some CPTP time dependent map $%
\varepsilon _{t}$. This external interaction of the coin is taken to have
been initiated at some past time $0,$ and to continuously happen in time
while atom crosses the cavity. Entering into the cavity at some time $t,$ in
state $\varepsilon _{t}(\rho _{c})$, the atom interacts instantaneously,
being the coin part of QRW, with the walker/cavity mode. For the $V^{2}$
model two such coin+walker interactions occur: one interaction at time $t,$
that changes their combine state as $\varepsilon _{t}(\rho _{c})\otimes \rho
_{w}\rightarrow V_{cl}(\varepsilon _{t}(\rho _{c})\otimes \rho
_{w})V_{cl}^{\dagger },$ and a second one later on at time $t+\tau ,$ that
effects the change of state: $V_{cl}(\varepsilon _{t}(\rho _{c})\otimes \rho
_{w})V_{cl}^{\dagger }$ $\rightarrow V_{cl}(\varepsilon _{\tau }\otimes
\mathbf{1(}V_{cl}(\varepsilon _{t}(\rho _{c})\otimes \rho
_{w})V_{cl}^{\dagger }))V_{cl}^{\dagger }.$ The $V_{cl}$ 's are realized by
sudden on-off switches of the mode+atom interaction, possibly by driving the
cavity mode off resonance. Then the two interactions taken together
constitute one step of the walk. Subsequently the atom is considered leaving
the cavity, the time clock is been reset, and a new atom is entering the
cavity.(A realistic cavity QED model for implementing QRW in atom+mode
interactions has been proposed in \cite{sanders}.) Explicitly the total
change of walker's density matrix between two successive steps is given by
the equation
\begin{equation}
\rho _{w}^{(n)}=\varepsilon _{V^{2}}(\rho _{w}^{(n-1)})=Tr_{c}\left[
AdV_{cl}\cdot \varepsilon _{\tau }\otimes \mathbf{1}\cdot AdV_{cl}\cdot
\varepsilon _{t}\otimes \mathbf{1}\cdot (\rho _{c}\otimes \rho _{w}^{(n-1)})%
\right] ,
\end{equation}%
for $n=1,2,...,$or more explicitly
\begin{equation}
\rho _{w}^{(n)}=\varepsilon _{V^{2}}(\rho _{w}^{(n-1)})=\sum_{ijk=\pm
1}\langle i|\varepsilon _{t}(\rho _{c})|j\rangle \langle k|\varepsilon
_{\tau }(|i\rangle \langle j|)|k\rangle E_{i+k}\rho _{w}^{(n-1)}E_{j+k}.
\end{equation}%
This shows that the $\varepsilon -$quantized $V^{2}$ walk proceeds with
steps of length $0$ and $2,$on the ladder of walker states, with weights
determined by the time dependent CPTP $\varepsilon _{t}.$This same map
actually serves as the source of quantization of classical $V_{cl}^{2}$
walk. It is important however to emphasize that in this walk the physical
origin of quantization is not an ad hoc imposed unitary rotation matrix in
coin space, as it has been in most cases following the $U-$quantization
rule, but instead it is the physical process of natural interaction of coin
system with some external agent. As an example we can consider the case of a
Rabi oscillating two-level atom that decays spontaneously. Such physical
conditions render the $\varepsilon -$generalization of quantization for a
classical walk a well motivated one.

The one-step map $\varepsilon _{V^{2}}$ previously introduced is determined
by parameters; these are some e.g $\lambda $ parameter measuring the
strength of $\varepsilon _{t}$, and the parameters $t$ and $\tau ,$
determining the time intervals of coin+walker interactions. Those time
parameters should be finely tuned, so that is possible for the two
interactions to take place during the time the atom/coin spends in the
cavity. This can be decided by selecting the velocity of the atomic beam
crossing the cavity.

To proceed with the problem of time evolution we introduce walker's density
matrix, $\rho _{w}=\int_{0}^{2\pi }\int_{0}^{2\pi }\rho (\phi ,\phi
^{^{\prime }})|\phi \rangle \langle \phi ^{^{\prime }}|d\phi d\phi
^{^{\prime }}.$ Due to linearity of evolution we only need to compute%
\begin{eqnarray}
\varepsilon _{V^{2}}(|\phi \rangle \langle \phi ^{^{\prime }}|)
&=&Tr_{c}\left( V_{cl}(\phi )\varepsilon _{\tau }\left( V_{cl}(\phi
)\varepsilon _{t}(\rho _{c})V_{cl}(\phi ^{^{\prime }})^{\dagger }\right)
V_{cl}(\phi ^{^{\prime }})^{\dagger }\right) |\phi \rangle \langle \phi
^{^{\prime }}|  \notag \\
&\equiv &A(\phi ,\phi ^{^{\prime }})|\phi \rangle \langle \phi ^{^{\prime
}}|,
\end{eqnarray}
then
\begin{equation}
\varepsilon _{V^{2}}^{n}(\rho _{w})=\int_{0}^{2\pi }\int_{0}^{2\pi }\rho
(\phi ,\phi ^{^{\prime }})A(\phi ,\phi ^{^{\prime }})^{n}|\phi \rangle
\langle \phi ^{^{\prime }}|d\phi d\phi ^{^{\prime }}.
\end{equation}
In formulas above matrix $V_{cl}(\phi ),$ is defined in the eigenbasis of
step operators as $V_{cl}=\int V_{cl}(\phi )|\phi \rangle \langle \phi
|d\phi ,$where $V_{cl}(\phi )=$diag$(e^{i\phi },e^{-i\phi }).$ Namely in $%
\phi $-basis we have that $E_{\pm }=e^{\pm i\Phi },$ where $\Phi =\int \phi
|\phi \rangle \langle \phi |d\phi ,$ is a Hermitean phase angle operator.

The discrete distribution determining the $m$-th site occupation probability
on the walker's ladder after $n$ steps is
\begin{equation}
P_{m}^{(n)}\equiv \langle m|\varepsilon _{V^{2}}^{n}(\rho _{w})|m\rangle =%
\frac{1}{(2\pi )^{2}}\int_{0}^{2\pi }\int_{0}^{2\pi }d\phi d\phi ^{^{\prime
}}A^{n}(\phi ,\phi ^{^{\prime }})e^{-im(\phi -\phi ^{^{\prime }})}.
\end{equation}%
Change to variables $\phi _{\pm }=\phi \pm \phi ^{^{\prime }},$ in the
preceding expression shows that \textit{if }$A(\phi _{+},\phi _{-}),$\textit{%
\ is independent from variable }$\phi _{+},$ then the $\phi _{+}$ integral
can be carried out and the probability becomes
\begin{equation}
P_{m}^{(n)}\equiv \langle m|\varepsilon _{V^{2}}^{n}(\rho _{w})|m\rangle =%
\frac{1}{2\pi }\int_{0}^{2\pi }d\phi _{-}\text{ }A^{n}(\phi _{-})e^{-im\phi
_{-}},
\end{equation}%
then the $A(\phi _{-})$ can be identified with the characteristic function
of the transition probability function of a classical random walk\cite%
{hughes}. This is a general result independent of quantization rule that can
be used as a \textit{criterion of classicality} of quantum walk. Let us give
some examples of classical walks quantized according to $U$ and $\varepsilon
$ rules, and impose on them the criterion of classicality.

Examples from $U-$quantization rule:

i) The $V$ model with evolution $\varepsilon _{V}(\rho
_{w}^{(n)})=Tr_{c}V\rho _{c}\otimes \rho _{w}^{(n-1)}V^{\dagger },$ gives $%
A_{V}(\phi _{+},\phi _{-})=\langle +|\rho _{c}|+\rangle e^{i\phi
_{-}}+\langle -|\rho _{c}|-\rangle e^{-i\phi _{-}}.$This fulfills the
classicality criterion and essentially leads to classical occupation
probabilities drawn from the diagonal elements of walker's density matrix.

\textit{ii}) The $V^{2}$ model with evolution $\varepsilon _{V}(\rho
_{w}^{(n)})=Tr_{c}V^{2}\rho _{c}\otimes \rho _{w}^{(n-1)}V^{2\dagger },$
initial $\rho _{c}=|+\rangle \langle +|,$ and reshuffling matrix $U_{\frac{%
\pi }{4}}=e^{i\frac{\pi }{4}\sigma _{2}}$ , been a $\frac{\pi }{4}$-rotation$%
.$ This gives \cite{ellsmyrn1}%
\begin{equation}
A_{V^{2}}(\phi _{+},\phi _{-})=\cos ^{2}\phi _{-}-i\cos \phi _{+}\sin \phi
_{-}.
\end{equation}

The asymptotic characteristic function (cf. eq.(\ref{HHH})), that determines
the limiting probabilities of the walk, is for this example

\begin{equation}
h(\phi )=-i\left[ \partial _{\phi }A(\phi ,\phi ^{\prime })\right] _{\phi
^{\prime }=\phi }=-\cos 2\phi .
\end{equation}

Examples from $\varepsilon -$quantization rule:

\textit{iii}) Assume initially we have a coin being in some mixed state $%
\rho _{c}=$diag$(q,1-q),\mathbf{\ \ }0\mathbf{\leq }q\leq 1,$ and that it
suffers spontaneous de-excitation of its upper state with rate $\lambda $,
then its state is described by the map \cite{nielsenchuang},

\begin{equation}
\varepsilon _{t}(\rho _{c})=S_{0}(t)\rho _{c}S_{0}^{\dagger
}(t)+S_{1}(t)\rho _{c}S_{1}^{\dagger }(t),
\end{equation}%
where $\cos (\lambda t)=\sqrt{1-e^{-2\lambda t}},$ and
\begin{equation}
S_{0}(t)=\left(
\begin{array}{cc}
\cos (\lambda t) & 0 \\
0 & 1%
\end{array}%
\right) ,\text{ }S_{1}(t)=\left(
\begin{array}{cc}
0 & 0 \\
\sin (\lambda t) & 0%
\end{array}%
\right) .
\end{equation}%
Such a coin enters the cavity in state $\varepsilon _{t}(\rho _{c})=$diag$%
(q\cos ^{2}\lambda t,-q\cos ^{2}\lambda t+1).$ Then we obtain that
\begin{equation}
A_{V^{2}}(\phi _{+},\phi _{-})=e^{-i2\phi _{-}}(1-q\cos ^{2}\lambda
t)+e^{i2\phi _{-}}q\cos ^{2}\lambda t\cos ^{2}\lambda \tau +q\cos
^{2}\lambda t\sin ^{2}\lambda \tau
\end{equation}%
For $q=\frac{1}{2},$ we have the initial density matrix $\rho _{c}=\frac{1}{2%
}\mathbf{1,}$ as special case. The criterion of classicality is also
fulfilled here, so the diagonal elements of evolved walker's density matrix
can be identified with a classical distribution.

\textit{Statement }: Direct calculation verifies the general statement that
classicality prevails in any $\varepsilon -$quantized $V^{k}$ model,
independently from the initial condition of coin system, as long as the
quantization proceeds by employing an $\varepsilon $ quantization map with
Kraus generators being matrices having only diagonal or only anti-diagonal
elements; such examples of $\varepsilon $ maps are e.g the $X,Y,Z,$ the
depolarization channels, the mentioned spontaneous emission channel, as well
as the transformation induced on the coin/atom after tracing out the bosonic
field freedom in Jaynes-Cummings model and in its various modifications
(c.f. \cite{ellsmyrn1,ellsmyrn2})$.$

\textit{iv}) \ If initially we have a coin being in some mixed state $\rho
_{c}=$diag$(q,1-q),\mathbf{\ \ }0\mathbf{\leq }q\leq 1,$ but $q\neq \frac{1}{
2},$ and if the quantization proceeds by using the map

\begin{equation}
\varepsilon (\rho _{c})=R_{0}\rho _{c}R_{0}^{\dagger }+R_{1}\rho
_{c}R_{1}^{\dagger },
\end{equation}%
where
\begin{equation}
R_{0}=\frac{1}{\sqrt{2}}\mathbf{1},\text{ }R_{1}=\frac{1}{\sqrt{2}}U_{\frac{%
\pi }{4}},
\end{equation}%
we then obtain that
\begin{equation}
A_{V^{2}}(\phi _{+},\phi _{-})=\frac{3(1+2q)}{16}e^{i2\phi _{-}}+\frac{%
3(1-2q)}{16}e^{-i2\phi _{-}}+\frac{i(1-2q)}{4}\sin \phi _{-}\cos \phi _{+}+%
\frac{1}{4}.  \label{alfa}
\end{equation}%
For $q=0,1,$ we have the initial density matrix $\rho _{c}=|-\rangle \langle
-|,|+\rangle \langle +|\mathbf{,}$ respectively. By comparing examples
\textit{ii}) and \textit{iv}) we see that we pass from the $U-$quantization
rule of the former to the $\varepsilon -$ quantization rule of the latter by
simply mixing the reshuffling unitary matrix $U_{\frac{\pi }{4}}$ with the
unit matrix. Such mixing is relevant in cases where there is probabilistic
uncertainty as to whether reshuffling matrix is applied or not.

The criterion of classicality is not satisfied here, therefore we have a
genuine $\varepsilon $-quantized random walk. By virtue of eq.(\ref{alfa}),
the asymptotic characteristic function is (c.f. eq.(\ref{HHH}))
\begin{equation}
h(\phi )=-\frac{3}{4}+\frac{1-2q}{4}\cos 2\phi .  \label{h_for_q_quant}
\end{equation}
Closing this section we note that both quantization rules give non trivial
models for quantum walks and that in some cases, as e.g in the last example
the $\varepsilon -$rule is presented as a necessary amendment of $U$-rule
when the latter can not be applied since the reshuffling matrix is either
not exactly known or is not accurately applied.

\section{Quantum Simulation of Asymptotics}

\textit{Asymptotics}: The dynamics of quantum walk can be described by the
quantum statistical moments of the observable of \textit{position operator} $%
\ L,$ evaluated e.g after $n$ steps. We obtain for its statistical moments
\begin{equation}
\langle L^{s}\rangle _{n}\equiv Tr(L^{s}\rho _{w}^{(n)})=\frac{1}{2\pi i^{s}}%
\int\limits_{0}^{2\pi }d\phi \left[ \partial _{\phi }^{s}\left[ \rho (\phi
,\phi ^{\prime })A^{n}(\phi ,\phi ^{\prime })\right] \right] _{\phi ^{\prime
}=\phi }.  \label{mom_def}
\end{equation}%
Next we study the asymptotic behavior of moments when the number of steps $n$
is large. \ In this case we have that
\begin{eqnarray}
\langle L^{s}\rangle _{n} &=&\frac{n^{s}}{2\pi i^{s}}\int\limits_{0}^{2\pi
}d\phi \rho (\phi ,\phi )\left[ \partial _{\phi }A(\phi ,\phi ^{\prime })%
\right] _{\mid \phi ^{\prime }=\phi }^{s}+O(n^{s-1})  \notag \\
&\equiv &\frac{n^{s}}{2\pi }\int\limits_{0}^{2\pi }d\phi \rho (\phi ,\phi
)h(\phi )^{s}+O(n^{s-1}).  \label{asmom1}
\end{eqnarray}%
Here we have introduced the \textit{asymptotic characteristic function}
(acf) $h(\phi ),$ of the walk as%
\begin{equation}
h(\phi )=-i\left[ \partial _{\phi }A(\phi ,\phi ^{\prime })\right] _{\phi
^{\prime }=\phi }.  \label{HHH}
\end{equation}%
\ \ \ \ \ The scaled by time limiting statistical moment of position
operator is then obtained to be
\begin{equation}
\left\langle \left( \frac{L}{n}\right) ^{s}\right\rangle _{\infty }\equiv
\frac{1}{2\pi }\int\limits_{0}^{2\pi }d\phi \rho (\phi ,\phi )h(\phi )^{s}.
\end{equation}

For the $U$-quantized $\varepsilon _{V^{k}}$ model since $h(\phi )=$Im$%
Tr_{c}(V^{k\dag }(\phi )V^{k}(\phi )\rho _{c}),$ as can be obtained by
elaborating on last equation, we have that
\begin{equation}
h(\phi )=Tr_{c}\left[ (\sigma +V^{\dagger }(\phi )\sigma V(\phi )+\cdots
+V^{\dagger k-1}(\phi )\sigma V^{k-1}(\phi ))\rho _{c}\right] ,  \label{HH}
\end{equation}%
where $\sigma =U^{\dagger }\sigma _{3}U$ is a rotated $\sigma _{3}$ Pauli
matrix.

It is important to note that the limiting positional moments provide all
necessary information for a sufficient understanding of \ the ensuing walker
asymptotic dynamics, and as seen from above these moments are expressed as
classical mean values of the powers of function $h(\phi )$ of the stochastic
variable $\phi ,$ that takes values around a circle with pdf $\frac{1}{2\pi }%
\rho (\phi ,\phi ).$ Hence we will seek the exact knowledge of function $%
h(\phi )$ next, by the technique of quantum simulating the system of walk.

\textit{Quantum Simulation}: The concept of quantum simulation of dynamics
or statistics of a quantum system by another quantum system constitute a
final goal for Quantum Information Science, since an alleged universal
quantum computer device would function as an efficient simulator of any
quantum process. In more modest claims a special purpose quantum system
could be constructed and set up to interact with its environment, so that
its dynamical or statistical behavior would simulate the respective dynamics
or statistics of a given quantum system. The simulator system is possibly
different from the original system, in e.g its dimension, type of
interactions or necessary physical and computational resources required from
its time evolution. However both original and simulator systems are both
governed by laws of quantum mechanics\cite{feynman,lloyd,zalka,soma}.

To construct a quantum simulation of asymptotic behavior of $U$-quantized
walk, we proceed by simulating quantum mechanically its asymptotic
characteristic function\textit{\ }$h(\phi ).$ Let us refer to eq.(\ref{HH}),
and introduce firstly the CPTP map \ \
\begin{equation}
\varepsilon _{\phi }^{\ast }(\sigma )=\frac{1}{k}(\sigma +V^{\dagger }(\phi
)\sigma V(\phi )+\cdots +V^{\dagger k-1}(\phi )\sigma V^{k-1}(\phi )),
\end{equation}%
then we cast \textit{acf} of eq.(\ref{HH}), in the form $h(\phi
)=kTr_{c}(\varepsilon _{\phi }^{\ast }(\sigma )\rho _{c}).$ Equivalently we
can express \textit{acf }in the form $h(\phi )=kTr_{c}(\varepsilon _{\phi
}(\rho _{c})\sigma ),$ where utilizing the cyclic property of trace, the
dual map $\varepsilon _{\phi }$ of the preceding map $\varepsilon _{\phi
}^{\ast },$ has been used that reads
\begin{equation}
\varepsilon _{\phi }(\rho _{c})=\frac{1}{k}\left( \rho _{c}+V(\phi )\rho
_{c}V^{\dagger }(\phi )+\cdots +V^{k-1}(\phi )\rho _{c}V^{\dagger k-1}(\phi
)\right) .  \label{ef_rho}
\end{equation}%
It is now possible to express the scaled $s$-th moment of \ quantum variable
$L,$ after the $n$-th step of the walk, in the suggestive form
\begin{equation}
\left\langle \left( \frac{L}{kn}\right) ^{s}\right\rangle
_{n}=\int\limits_{0}^{2\pi }\frac{\rho (\phi ,\phi )d\phi }{2\pi }%
(Tr_{c}[\sigma \varepsilon _{\phi }(\rho _{c})])^{s}+O(n^{-1}).
\label{smoment}
\end{equation}%
Let us first elaborate on the first moment taken for $s=1$ in last equation;
to this end we introduce the $\phi -$average of the transformed density
matrix $\varepsilon _{\phi }(\rho _{c}),$ with respect to the probability
distribution function (pdf), $(\frac{\rho (\phi ,\phi )}{2\pi },$ $0<\phi
\leq 2\pi ),$ that reads
\begin{equation}
\overline{\varepsilon }(\rho _{c})=\int\limits_{0}^{2\pi }\frac{\rho (\phi
,\phi )d\phi }{2\pi }\varepsilon _{\phi }(\rho _{c}).  \label{e_bar}
\end{equation}%
Then we obtain the first moment for $n>>1,$ in the form
\begin{equation}
\left\langle \left( \frac{L}{kn}\right) \right\rangle _{\infty }=Tr_{c}(%
\overline{\varepsilon }(\rho _{c})\sigma )=\lim_{n\rightarrow \infty }\frac{1%
}{kn}Tr_{w}(\rho _{w}^{(n)}L).  \label{mom1}
\end{equation}
This is interpreted as saying that (c.f. first eq. above) in the asymptotic
regime of the walk, the expectation value of scaled variable $L$ is
proportional to the expectation value of observable $\sigma ,$ evaluated
with the coin density matrix been transformed by the $\phi -$average of $%
\varepsilon _{\phi }.$ In the second equation above, it is emphasized that
moment $\left\langle \left( \frac{L}{n}\right) \right\rangle _{\infty }$ is
initially defined as expectation value of walker's space observable in the
limit of large number of steps.

Next we continue our elaboration with the case for higher moments \ i.e $%
s>1. $ Referring to eq.(\ref{smoment}), the integrand $(Tr_{c}[\sigma
\varepsilon _{\phi }(\rho _{c})])^{s},$ by means of the property of trace $%
Tr(A\otimes B)=TrA\cdot TrB,$ or $TrA^{\otimes n}=(TrA)^{n},$ is expressed
as $(Tr_{c}[\sigma \varepsilon _{\phi }(\rho _{c})])^{s}=Tr_{c}[\sigma
\varepsilon _{\phi }(\rho _{c})\otimes \sigma \varepsilon _{\phi }(\rho
_{c})...\otimes \sigma \varepsilon _{\phi }(\rho _{c})].$ Further use of the
property $A\otimes B\cdot C\otimes D=AC\otimes BD,$ (dot sign emphasizes
ordinary matrix product), and of notation $A^{\otimes n}=A\otimes A\otimes
...\otimes A,$ for $n$-fold tensor product, yields that
\begin{equation}
(Tr_{c}[\sigma \varepsilon _{\phi }(\rho _{c})])^{s}=Tr_{c}[\sigma ^{\otimes
s}\cdot \varepsilon _{\phi }(\rho _{c})\otimes \varepsilon _{\phi }(\rho
_{c})\otimes ...\otimes \varepsilon _{\phi }(\rho _{c})].  \label{tensor_s}
\end{equation}%
We also need to introduce the $\phi -$average of the products of asymptotic
density matrix $\varepsilon _{\phi }(\rho _{c}),$ appearing above i.e
\begin{equation}
\overline{\varepsilon ^{s}}(\rho _{c}^{\otimes s})=\int\limits_{0}^{2\pi }%
\frac{\rho (\phi ,\phi )d\phi }{2\pi }[\varepsilon _{\phi }(\rho
_{c})\otimes \varepsilon _{\phi }(\rho _{c})\otimes ...\otimes \varepsilon
_{\phi }(\rho _{c})].  \label{tensor_e}
\end{equation}%
Combining eqs. \bigskip (\ref{tensor_s},\ref{tensor_e}), we cast eq.(\ref%
{smoment}) of the asymptotic $s$-th moment in following form%
\begin{equation}
\left\langle \left( \frac{L}{kn}\right) ^{s}\right\rangle _{\infty
}=Tr_{c}[\sigma ^{\otimes s}\overline{\varepsilon ^{s}}(\rho _{c}^{\otimes
s})]=\lim_{n\rightarrow \infty }\frac{1}{(kn)^{s}}Tr(\rho _{w}^{(n)}L^{s})
\label{moms}
\end{equation}%
Preceding equation (\ref{moms}) and its $s=1$ version in eq.(\ref{mom1}),
provide a framework for a quantum simulation of asymptotic statistics of
quantum walk on a line. Such a framework identifies the walker system of a
QRW with the system that is simulated by a second quantum system the
simulator which in this case can be identified with the coin system of the
QRW in question.

Indeed let us assume that we set up some appropriate Hamiltonian dynamics
for a composite system comprised by a two-level atom identified with the
coin system, and an ancillary system, so that after decoupling the two, by
tracing out the ancilla system, the coins finds themselves in state $%
\overline{\varepsilon ^{s}}(\rho _{c}).$ Then the quantum mean value of the
observable $\sigma ^{\otimes s},$ i.e. $\left\langle \sigma ^{\otimes
s}\right\rangle =Tr\sigma ^{\otimes s}\overline{\varepsilon ^{s}}(\rho
_{c}), $ will be equal with the asymptotic $s$-th moment, of the walker
system of the simulated QRW, by virtue of eq.(\ref{moms}). \bigskip The
problem of constructing appropriate Hamiltonian dynamics has no unique
solution, so next we provide a solution for the simplest nontrivial case of $%
k=2.$ For this case eq.(\ref{ef_rho}), becomes
\begin{equation}
\varepsilon _{\phi }(\rho _{c})=\frac{1}{2}\left( \rho _{c}+V(\phi )\rho
_{c}V^{\dagger }(\phi )\right) .
\end{equation}%
To unitarize this transformation of coin density matrix we introduce a 2D
auxiliary system, where $\rho _{a}=|+\rangle \langle +|$ is taken to be its
density matrix. Then we find that
\begin{equation}
\varepsilon _{\phi }(\rho _{c})=Tr_{a}W(\phi )(\rho _{a}\otimes \rho
_{c})W(\phi )^{\dagger },  \label{w_fi}
\end{equation}%
where the unitary matrix $W(\phi ),$ is chosen to be
\begin{equation}
W(\phi )=\frac{1}{\sqrt{2}}(\mathbf{1\otimes 1+}|-\rangle \langle +|\otimes
V(\phi )-|+\rangle \langle -|\otimes V(\phi )^{\dagger })=\frac{1}{\sqrt{2}}%
\left(
\begin{array}{cc}
\mathbf{1} & V(\phi )^{\dagger } \\
-V(\phi ) & \mathbf{1}%
\end{array}%
\right) .
\end{equation}%
Writing $W(\phi )=\exp H(\phi ),$ the associated Hamiltonian matrix reads $%
H(\phi )=\frac{\pi }{4}[\sigma _{+}\otimes V(\phi )^{\dagger }-\sigma
_{\_}\otimes V(\phi )],$ so that if we choose the reshuffling matrix to be a
$\frac{\pi }{4}$-rotation matrix, i.e. $V(\phi )=V_{cl}(\phi )U=e^{i\phi
\sigma _{3}}U_{\frac{\pi }{4}}=e^{i\phi \sigma _{3}}e^{i\frac{\pi }{4}\sigma
_{2}},$ then the Hamiltonian suggests a coupling of two spins and becomes%
\begin{equation}
H(\phi )=\frac{\pi }{4}[\sigma _{+}\otimes e^{-i\frac{\pi }{4}\sigma
_{2}}e^{-i\phi \sigma _{3}}-\sigma _{\_}\otimes e^{i\phi \sigma _{3}}e^{i%
\frac{\pi }{4}\sigma _{2}}].
\end{equation}%
Having constructed the Hamiltonian interaction that yields density matrix $%
\varepsilon _{\phi }(\rho _{c}),$ the $\phi -$averaged matrix $\overline{%
\varepsilon ^{s}}(\rho _{c}),$ written in the form
\begin{equation}
\overline{\varepsilon ^{s}}(\rho _{c})=Tr_{a}\int\limits_{0}^{2\pi }\frac{%
\rho (\phi ,\phi )d\phi }{2\pi }\left[ W(\phi )^{\otimes s}\left( \rho
_{a}\otimes \rho _{c}\right) ^{\otimes s}W(\phi )^{\otimes s\dagger }\right]
.  \label{e_bar_s}
\end{equation}%
requires the $s$-fold product unitary $W(\phi )^{\otimes s}=e^{\Delta H},$
with associated Hamiltonian the following multiple spin matrix $\Delta
H=H\otimes \mathbf{1}\otimes ...\otimes \mathbf{1+...+1}\otimes \mathbf{...1}%
\otimes H.$ Explicitly it involves $2s$ \ two-level systems coupled in
nearest neighbor form,
\begin{eqnarray}
\Delta H &=&\frac{\pi }{4}[\sigma _{+}\otimes V(\phi )^{\dagger }-\sigma
_{\_}\otimes V(\phi )]\otimes \mathbf{1}\otimes ...\otimes \mathbf{1}+\frac{%
\pi }{4}\mathbf{1}\otimes \lbrack \sigma _{+}\otimes V(\phi )^{\dagger
}-\sigma _{\_}\otimes V(\phi )]\otimes \mathbf{1}\otimes ...\otimes \mathbf{1%
}  \notag \\
&&\mathbf{+}\frac{\pi }{4}\mathbf{1}\otimes \mathbf{1}\otimes \mathbf{...}%
\otimes \lbrack \sigma _{+}\otimes V(\phi )^{\dagger }-\sigma _{\_}\otimes
V(\phi )]\mathbf{.}
\end{eqnarray}%
For example if initially $\rho _{w}=|0\rangle \langle 0|,$ i.e. $\rho (\phi
,\phi )=1,$ then the density matrix of eq.(\ref{e_bar_s}) is obtained by a
uniform average of unitary similarity transformation generated by the
Hamiltonian of last equation. Having constructed $\overline{\varepsilon ^{s}}%
(\rho _{c}),$ by means of the evolution of \ simulator system$,$ the moments
of quantum walker's system are obtained by averaging measurements of
observable $\sigma ^{\otimes s},$ in accordance with eq.(\ref{moms}).

\textit{Classical Stochastic Simulation }: An alternative way to implement
physically the transformation outlined in eq.(\ref{e_bar}), \ is to consider
the ensemble average of an appropriate stochastic unitary rotation acting on
a two-level system. In more concrete terms, let us consider the angular
random variable (rv) $\phi ,$distributed on circle by the pdf $\phi
\thicksim (\frac{\rho (\phi ,\phi )}{2\pi },$ $0<\phi \leq 2\pi ),$ and the
independent discrete rv $\nu ,$ uniformly distributed over the first $k$
natural numbers i.e. $\nu \thicksim (\{0,1,...,k-1\}).$ Then we form a
transformation of the coin density matrix as follows $\rho _{c}\rightarrow
V(\phi )^{\nu }\rho _{c}V(\phi )^{\dagger \nu }.$ This is a random
similarity transformation of the density matrix with the random unitary
matrix $V(\phi )^{\nu };$ its randomness is both due to angle $\phi $ and
due to exponent $\nu .$ Taking the statistical average of this
transformation over its two rv's with respect to their corresponding pdf's
to be denoted by $\left\langle V(\phi )^{\nu }\rho _{c}V(\phi )^{\dagger \nu
}\right\rangle _{\phi ,\nu },$ we write that
\begin{equation}
\left\langle V(\phi )^{\nu }\rho _{c}V(\phi )^{\dagger \nu }\right\rangle
_{\phi ,\nu }=\frac{1}{k}\int\limits_{0}^{2\pi }\frac{\rho (x,x)dx}{2\pi }%
\left( \sum_{m=0}^{k-1}V(x)^{m}\rho _{c}V(x)^{\dagger m}\right) \equiv
\overline{\varepsilon }(\rho _{c}).
\end{equation}
This is identical with eq.(\ref{e_bar}); note also that due to the
statistical independence of $\phi $ and $\nu $ variables, the double
statistical mean of rotations is obtained by evaluating successively the
mean for each variable i.e. $\overline{\varepsilon }(\rho _{c})=\left\langle
V(\phi )^{\nu }\rho _{c}V(\phi )^{\dagger \nu }\right\rangle _{\phi ,\nu
}=\left\langle \left\langle V(\phi )^{\nu }\rho _{c}V(\phi )^{\dagger \nu
}\right\rangle _{\phi }\right\rangle _{\nu }=\left\langle \left\langle
V(\phi )^{\nu }\rho _{c}V(\phi )^{\dagger \nu }\right\rangle _{\nu
}\right\rangle _{\phi }.$

This same implementation idea is further generalized to get the analogue of
eq.(\ref{e_bar_s}). For that $s$ identical two-level systems are needed
together with $s$ identical and statistically independent discrete random
variables, uniformly distributed over the first $k$ natural numbers i.e. $%
\nu _{i}\thicksim (\{0,1,...,k-1\}),$ $i=1,2,...,s,$ as well as an
independent circular rv $\phi \thicksim (\frac{\rho (\phi ,\phi )}{2\pi },$ $%
0<\phi \leq 2\pi )$. First we form the statistical average over the $s$
independent discrete variables
\begin{equation}
\varepsilon _{\phi }(\rho _{c})\otimes ...\otimes \varepsilon _{\phi }(\rho
_{c})=\left\langle V(\phi )^{\nu }\rho _{c}V(\phi )^{\dagger \nu
}\right\rangle _{\nu }\otimes ...\otimes \left\langle V(\phi )^{\nu }\rho
_{c}V(\phi )^{\dagger \nu }\right\rangle _{\nu }.
\end{equation}
Then we treat the resulting $s$-fold tensor product of density matrices $%
\varepsilon _{\phi }(\rho _{c}),$ as $\phi $ correlated matrix-valued random
variables, and consider their statistical average
\begin{equation}
\left\langle \varepsilon _{\phi }(\rho _{c})\otimes ...\otimes \varepsilon
_{\phi }(\rho _{c})\right\rangle _{\phi }=\int\limits_{0}^{2\pi }\frac{\rho
(\phi ,\phi )d\phi }{2\pi }\left\langle V(\phi )^{\nu }\rho _{c}V(\phi
)^{\dagger \nu }\right\rangle _{\nu }\otimes ...\otimes \left\langle V(\phi
)^{\nu }\rho _{c}V(\phi )^{\dagger \nu }\right\rangle _{\nu }.
\end{equation}
The doubly averaged density matrix is identical to $\overline{ \varepsilon
^{s}}(\rho _{c}),$ i.e. $\overline{\varepsilon ^{s}}(\rho _{c})=\left\langle
\varepsilon _{\phi }(\rho _{c})\otimes ...\otimes \varepsilon _{\phi }(\rho
_{c})\right\rangle _{\phi }.$

We note finally that the range $k$ of discrete rv $\nu $ is determined by
the kind of $V^{k}$ model of QRW the statistical moments of which we
simulate, and that similarly the number $s$ of two-level atoms involved in
simulation is determined by the order of quantum moment of walker's system
we intend to simulate. Two-level atoms, and the $\phi ,$ $\nu ,$ classical
stochastic variables are quantum and classical resources required for this
\textit{on the average} stochastic simulation of QRW. Note also that, the
main difference between quantum and stochastic simulations is in the way the
CPTP maps of coin systems are derived: in the stochastic case many runs of
random rotations are required \ so that a classical ensemble average is
formed, while in quantal case the prescribed total unitary evolution is
generated in one run and then an coin unconditional measurement provides the
final density matrix (c.f. \ref{w_fi}-\ref{e_bar_s}).

As to the experimental realization of the proposed stochastic implementation
we note that the transformation $\rho _{c}\rightarrow V(\phi )^{\nu }\rho
_{c}V(\phi )^{\dagger \nu },$ requires a random rotation $V(\phi )^{\nu
}=(V_{cl}(\phi )U)^{\nu }=(e^{i\phi \sigma _{3}}e^{i\frac{\pi }{4}\sigma
_{2}})^{\nu }.$ For the case of \ $k=2,$ i.e. $\nu =0,1,$ we need to flip
randomly between the two rotations $(\mathbf{1,}e^{i\phi \sigma _{3}}e^{i%
\frac{\pi }{4}\sigma _{2}}),$ of the coin system. Stochastic unitary
rotations of a two-level atom can experimentally be achieved by randomly
pulsating laser fields with appropriate fluctuating phases and intensities
\cite{stenholm}.

As corroboration of our theoretical results about quantum simulation, we
next provide numerical evaluation of the asymptotic probability density
function of the walker. This is done for the example of $U-$quantization
(c.f. eq.( \ref{HH}) for the models with $k=2,3$), and the example of $%
\varepsilon -$quantization (c.f. eq.(\ref{h_for_q_quant})). If $Y=h(\phi ),$
then the cumulative probability function in the interval $[y_{1},y_{2}]$
becomes
\begin{equation}
P(y_{1}\leq Y\leq y_{2})=\frac{1}{2\pi }\int\limits_{y_{1}\leq h(\phi )\leq
y_{2}}\rho (\phi ,\phi )d\phi .
\end{equation}%
In the application we choose $\rho (\phi ,\phi )=1,$corresponding to $\rho
_{w}=|0\rangle \langle 0|.$ In the numerical simulation to determine the
values of the probability density function over some small interval $%
I=[a,b], $ we count the number of times the random variable $\phi ,$ is such
that $y_{1}\leq h(\phi )\leq y_{2},$ and compute the fraction of the number
of successful counts by the overall number of counts. The results displayed
in fig. 1a, for the model $V^{2},$ quantized by the $U$-rule, show the well
known by now double-horn distribution (c.f. previous works cited in the
introductory chapter). As to the results of fig.1b and fig.2, they are new,
and provide information about the asymptotic behavior of the $k=3$ model
quantized by $U-$rule, and of the model $k=2,$ quantized according to $%
\varepsilon -$quantization rule. The results show some similarity in the
form of the distributions, but also important differences in the
displacements and the cutoffs of their supports. These should be detectable
features in realistic simulations of a walk with quantum systems. Finally,
these results should be compared and constrasted with those of \cite{brun}.
In that work the delayed tracing scheme with time step evolution maps $%
\varepsilon _{V},$\ $\varepsilon _{V^{2}},$\ $\varepsilon _{V^{3}},...,$\
\cite{ellinas1}, was used for QRW on line, in which coin decoherence has
been introduced, and modifies the quantum walk along the lines of $%
\varepsilon -$quantization rule. These results show destruction of quantum
features of the walk for increasing strength of decoherence. On the contrary
results indicated in fig. 2, show that for the analogous $\varepsilon -$%
quantized $\varepsilon _{V^{2}}$\ model with $\varepsilon _{V^{2}}^{n},$\ $%
n=1,2,...,$\ time step evolution maps, the quantum features are retained in
the long time regime, albeit in a modified form.

\section{Conclusions}

Quantization rules provide a broad and systematic framework for quantizing
classical random walks, taking into account phenomena such as losses,
decoherence, noise or coherent dynamics occurring in the coin systems.
Optical processes that may cause or on purpose induce similar phenomena in
coins may therefore constitute physical probes for the study of novel
features in quantum walks. The prospects of such quantum optical walks are
further enhanced by the possibility of using the coin systems not only as a
trigger of the walk, but also as a quantum simulator of its dynamics and
statistics, especially of its long terms characteristics \cite{dirac}. In
this framework phenomena related to quantum-classical transitions in
walker's evolution, and to quantum coin+walker entanglement, especially in
the asymptotic regime of a walk, find new theoretical and experimental
possibilities for a in-depth investigation. To some of these topics we aim
to return elsewhere.

\vskip0.5cm Acknowledgments: This work was supported by "Pythagoras II" of
EPEAEK research programme.

\pagebreak

Figure captions

Figure 1a,1b

Asymptotic probability density function, for walker's scaled position
variable. It refers to the $V^{2}$ model in fig. 1a (the $V^{3}$ model in
fig. 1b), quantized by the $U$-rule, with reshuffling matrix the $\frac{\pi
}{4}$ rotation matrix, and initial coin chosen in excited state.

Figure 2

Asymptotic probability density function, for walker's scaled position
variable. It refers to the $V^{2}$ model, quantized by the $\varepsilon $%
-rule, with initial coin chosen in excited state.


\begin{thebibliography}{99}
\bibitem{vk} These model have been introduced in \cite{ellinas1},
generalized in \cite{ellinas}, and studied extensively in \ \ \ \ \cite%
{ellsmyrn1,ellsmyrn2}.
\end{thebibliography}
\end{document}